\documentclass[12pt]{article}
\usepackage{amssymb,amsmath}
\begin{document}


\sloppy
\title
{\Large Equivalent sets of solutions of the Klein-Gordon equation with
 a constant electric field}

\author
 {
       A.I.Nikishov
          \thanks
             {E-mail: nikishov@lpi.ru}
  \\
               {\small \phantom{uuu}}
  \\
           {\it {\small} I.E.Tamm Department of Theoretical Physics,}
  \\
               {\it {\small} P.N.Lebedev Physical Institute, Moscow, Russia}
  \\
 }
%
\maketitle
        \begin{abstract}

In connection with the problem of choosing the in- and out-states among
the solutions of a wave equation with one-dimensional potential we study
nonstationary and "stationary" families of complete sets. A nonstationary
set consists of the solutions with the quantum number $p_v=p^0v-p_3.$
It can be obtained from the nonstationary set with quantum number $p_3$ by a
 boost along $x_3$-axis (along the direction of the electric field) with
velocity $-v$. By changing the gauge the solutions in all sets  can be
brought to one and the same potential without changing quantum numbers.
Then the transformations of solutions of one set (with quantum number
$p_v$) to the  solutions in another set (with quantum number $p_{v'}$) have
the group properties.

The "stationary" solutions and sets possess the same properties as the
nonstationary ones and are
obtainable from stationary solutions with quantum number $p^0$ by the same
boost.

It turns out that any set can be obtained from any other by gauge
manipulations. So all sets are equivalent and the classification
(i.e. ascribing the frequency sign and in-, out- indexes) in any set is
determined by the classification in $p_3$-set , where it is evident.

\end{abstract}
\section{Introduction and statement of the problem}
The choice of in- and out- solutions in [1-2] (see also [3]) disagrees with
 that in [4-5].
So the classification problem in different sets of solutions is on hand.
We show that the choice of gauge of potential, describing the considered
field, fixes the natural quantum number for this gauge. So each gauge define
a complete set of solutions. Then we can go over to any other gauge without
changing the quantum number. In this way we can relate solutions in different
sets characterized by different quantum numbers. But changing gauge does not
change neither the physical system nor the classification in the sets. These
simple considerations open the way to the solution of the stated problem,
because the classification in the set with quantum number $p_3$ is evident.

Already in classical mechanics we can see in what sense fixing the gauge
fixes the conserved number. The solutions of the classical equation of motion
 for a particle in a constant electric field have the form
$$
\pi^0(t)=p_3\sinh\varepsilon s+p^0\cosh\varepsilon s,\quad \varepsilon=
\frac{eE}{m},
$$
  $$
\pi_3(t)=p_3\cosh\varepsilon s+p^0\sinh\varepsilon s,
  $$
$$
eEt=p_3(\cosh\varepsilon s-1)+p^0\sinh\varepsilon s, \eqno(1)
$$
  $$
eEx_3=p_3\sinh\varepsilon s+p^0(\cosh\varepsilon s-1).
  $$
We use the metric
$$
\eta_{\mu\nu}={\rm diag}(-1,1,1,1). \eqno(2)
$$
The motion in the direction perpendicular to the field remains free and we are
not interested in it here.

At first we take the vector-potential in the form
$$
A_{\mu}=-\delta_{\mu3}Ex_v,\quad x_v=t-vx_3,\quad0\le v<0.\eqno(3)
$$
It follows from (1) and (3) that
$$
\pi^0v-\pi_3+e(vA^0-A_3)=p^0v-p_3\equiv p_v={\rm Const}.  \eqno(4)
$$
(Vector-potential is taken on particle's trajectory)
 The potential
$$
A_3=-\frac{Ex_v}{1-v^2},\quad A^0=-\frac{vEx_v}{1-v^2}, \eqno(5)
$$
gives the same conserved number. This potential is
obtainable from (3) at $v=0$ by the above mentioned boost; the electric field
remains unchanged.

Similarly, for the potential
$$
A^0=-Ex_s,\quad x_s=x_3-st,\quad 0\le s<1,     \eqno(6)
$$
we find
$$
\pi^0-s\pi_3+e(A^0-sA_3)=p^0-sp_3\equiv p_s={\rm Const},    \eqno(7)
$$
and the same for the potential
$$
  A_3=-\frac{sEx_s}{1-s^2}.\quad A^0=-\frac{Ex_s}{1-s^2}, \eqno(8)
$$
obtainable from (6) at $s=0$ by the boost along $x_3$ with velocity $-s$.
We might denote $s$ by $v$  as before, but we prefer to have separate
notation for a different situation.

Now we consider the Klein-Gordon equation
$$
\partial_{\mu}\partial^{\mu}\psi=[2ieA^{\mu}\partial_{\mu}+
ie(\partial_{\mu}A^{\mu})+e^2A^2+m^2]\psi.                            \eqno(9)
$$
The vector-potential, describing the constant electric field, is taken
 at first in the general form
$$
A^{\mu}=a^{\mu}\varphi,\quad \varphi=k\cdot x,\quad k^{\mu}=(k^0,0,0,k_3),\quad
a^{\mu}=(a^0,0,0,a_3).\eqno(10)
$$
The solution for (9) is sought in the form
$$
\psi_p=C_p\exp\{i[p\cdot x-\frac{p\cdot k}{k^2}\varphi+
\frac{ea\cdot k}{2k^2}\varphi^2]\}v_p.                    \eqno(11)
$$
Substituting this in (9) gives
$$
[\frac{d^2}{d\varphi^2}+\left(\frac{p\cdot k}{k^2}\right)^2+
2c_1\varphi+c_2\varphi^2]v_p=0,
                                                             \eqno(12)
$$
  $$
c_1=\frac{ea\cdot p}{k^2}-\frac{p\cdot k ea\cdot k}{(k^2)^2},\quad
c_2=\left(\frac{ea\cdot k}{k^2}\right)^2-\frac{e^2a^2}{k^2}=
\frac{e^2E^2}{(k^2)^2}.
                                              \eqno(13)
  $$
Going over to the variable
$$
\tau=-c_2^{\frac14}(\varphi+\frac{c_1}{c_2})    \eqno(14)
$$
reduces (12) to the equation for the parabolic cylinder function [6]
$$
[\frac{d^2}{d\tau^2}+\tau^2+\tilde\lambda]v_p=0,\quad \tilde\lambda=
-\lambda  {\rm sign}k^2,\quad\lambda=\frac{m^2+p_1^2+p_2^2}{|eE|}.   \eqno(15)
$$

In contrast to [4, 7-8] we assume in this paper that the charge of a scalar
particle $e=-|e|$. Now we consider two separate families  of potential (10).

\section{Nonstationary solutions}

We specify (10) as follows
$$
a_{\mu}=\delta_{\mu3}a,\; k^{\mu}=(\omega,0,0,v\omega),\;
A_3=-Ex_v, \; a\omega=E,\; x_v=t-vx_3,\; k^2=\omega^2(v^2-1).\eqno(16)
$$
The conserved quantum number is $p_v=p^0v-p_3$, cf (4). It follows from (13)
and (14) that
$$
\tau_v=\frac{\pi_v}{\sqrt{|eE|(1-v^2)}},\quad \pi_v=p_v-eEx_v,\quad
p_v=p^0v-p_3. \eqno(17)
$$
For brevity reasons we drop the dependence of the wave function on $x_1, x_2$
(i.e. we drop the factor $\exp[ip_1x_1+ip_2x_2]$). Then for the phase in (11)
we have
$$
p_3x_3-p^0t-\frac{k\cdot p}{k^2}\varphi+\frac{ea\cdot k}{2k^2}\varphi^2=
\frac{p_v(tv-x_3)}{1-v^2}-\frac{eEvx_v^2}{2(1-v^2)}.                \eqno(18)
$$
In this Section we compare solutions bringing them to the potential
$$
\tilde A_{\mu}=-\delta_{\mu3}Et=A_{\mu}+\frac{\partial\eta}{\partial x^{\mu}}.
  \eqno(19)
$$
Taking into account $A_{\mu}$ in (16) we find
$$
\eta=-\frac{Evx_3^2}2, \quad \psi_{p_v}(x|\tilde A)=e^{ie\eta}\psi_{p_v}(x|A).
    \eqno(20)
$$
So instead of (18) we have
$$
\vartheta_v(x|\tilde A)\equiv\vartheta_v=\frac{p_v(tv-x_3)}{1-v^2}-
\frac{eEvx_v^2}{2(1-v^2)}-\frac{eEvx_3^2}2.  \eqno(21)
$$

For uniformity we denote the wave function with quantum number $p_v$ as
$\psi_{p_v}$. Yet it should be remembered that for $v\to0$  this function
goes over into $\psi_{p_3,}$, not into $\psi_{-p_3}$ as one might think
looking at the definition of $p_v$ in (17).

Now we are in a position to write down and classify
 $\psi_{p_v}\equiv\psi_{p_v}(x|\tilde A)$-solutions. We put the frequency sign
(see [4]) before $\psi$-function in the lower position for in- solution and
 in the
upper position for out- solution. The in- (out-)solution has only one wave
of indicated frequency for $t\to-\infty (t\to\infty)$. So
$$
{}_{\pm}\psi_{p_v}=C_{pv}e^{i\vartheta_v}
D_{\pm i\frac{\lambda}{2}-\frac12}(-e^{\mp\frac{\pi}{4}}T_v),\;
{}^{\pm}\psi_{p_v}=C_{pv}e^{i\vartheta_v}
D_{\mp i\frac{\lambda}{2}-\frac12}(e^{\pm i\frac{\pi}{4}}T_v),
$$
  $$
T_v=\sqrt2\tau_v=\sqrt{\frac{2}{|eE|(1-v^2)}}(p_v-eEx_v),\;
C_{pv}=[2|eE|(1-v^2)]^{-\frac14}e^{-\frac{\pi\lambda}{8}}.     \eqno(22)
$$
${}_{\pm}\psi_{p_v}$ are normalized as follows
$$
\int_{-\infty}^{\infty} dx_3{}_{\pm}\psi_{p'_v}^
*i\buildrel\leftrightarrow\over{\frac d{dt}}
{}_{\pm}\psi_{p_v}=\pm2\pi\delta(p'_v-p_v) \eqno(23)
$$
and similarly for ${}^{\pm}\psi_{p_v}$. The classification is called forth
by the condition that for $v\to0$ $\psi_{p_v}$ goes over into $\psi_{p_3}$
and the classification of the latter functions is substantiated in [4].

From the relations between the parabolic cylinder functions it follows
$$
{}_+\psi_{p_v}=c_{1p}\,{}^+\psi_{p_v}+c_{2p}\,{}^+\psi_{p_v},
$$
  $$
{}_-\psi_{p_v}=c_{2p}^*\,{}^+\psi_{p_v}+c_{1p}^*\,{}^-\psi_{p_b}   \eqno(24)
   $$
 The Bogoliubov coefficients $c_{1p}, c_{2p}$ depend only on $\lambda$, i.e.
only on $p_{\perp}^2=p_1^2+p_2^2$ [4]:
$$
c_{1p}=\sqrt{2\pi}\Gamma^{-1}(\frac{1-i\lambda}{2})
\exp[-\frac{\pi}{4}(\lambda-i)],\;c_{2p}=\exp[-\frac{\pi}{2}(\lambda+i)],
\:|c_{1p}|^2-|c_{2p}|^2=1. \eqno(25)
$$

As shown in [7-8], $\psi_{p_v}$ can be obtained from $\psi_{p_3}$ by a boost
along $x_3-$axis with velocity $-v$ with subsequent regauging to the
potential $\tilde A_{\mu}$ in (19).

Now we consider the limiting case $v\to1$ and obtain the set of functions
with quantum number $p^-=p^0-p_3$ (a set in a system "with infinite momentum").
As seen from (17)
$$
\left.\tau_v\right|_{v\to1}\to\infty\, {\rm sign \pi_v,}\quad \pi_v\to\pi^-=
p^--eEx^- \eqno(26)
$$
and we need the asymptotic expansions of the parabolic cylinder functions.
 These expansions
 contain factors $\exp[\pm i\tau^2/2]$. For example, we have
$$
\left.D_{\frac{i\lambda}{2}-\frac12}[-(1-i)\tau_v]\right|_{\tau_v\to-\infty}
\to(2\tau_v^2)^{-\frac14}\exp[\frac{\pi}{8}(\lambda+i)+i\frac{\lambda}{4}\ln2
+\frac{i\tau_v^2}{2}+\frac{i\lambda}{2}\ln(-\tau_v)],\eqno(27)
$$
  $$
\left.D_{\frac{i\lambda}{2}-\frac12}[-(1-i)\tau_v]\right|_{\tau_v\to\infty}
\to\sqrt{2\pi}\Gamma^{-1}(\frac{1-i\lambda}{2})
(2\tau_v^2)^{-\frac14}\exp[-\frac{\pi}{8}(\lambda-i)-i\frac{\lambda}{4}\ln2
-\frac{i\tau_v^2}{2}-\frac{i\lambda}{2}\ln(\tau_v)]+
  $$
$$
 (2\tau_v^2)^{-\frac14}\exp[-\frac{3\pi}{8}(\lambda+i)+i\frac{\lambda}{4}\ln2
+\frac{i\tau_v^2}{2}+\frac{i\lambda}{2}\ln(\tau_v)].\eqno(28)
$$
Therefore the asymptotic expressions of $\psi-$functions in (22) contain the factors
$\exp[i(\vartheta_v\pm\frac{\tau_v^2}{2})]$. Simple calculations give
$$
\vartheta_v\pm\frac{\tau_v^2}{2}=\pm\frac{p_v^2}{2|eE|(1-v^2)}
\pm\frac{|eE|x_v^2}{2(1\mp v)}\pm\frac{p_vx^{\mp}}{1\mp v}+\frac{|eE|vx_3^2}{2},
\;x^{\pm}=t\pm x_3.\eqno(29)
$$

If we form a wave packet using functions with phases (29), then only the
lower sign case contributes in the limit $v\to1$. The divergence of the
constant term on the right-hand side of (29) is not dangerous and can be
removed by the compensating term in the phase of the weight function of
 the packet. Hence in the limit $v\to1$ terms with phase
$(\vartheta_v+\tau_v^2/2)$ may be dropped.

It follows from (22) and (27-28) that
$$
\left.{}_+\psi_{p_v}(x|\tilde A)\right|_{v\to1}\to\exp[-\frac{i\pi}{8}-
\frac{i\lambda}{4}\ln\frac{2}{1-v^2}-\frac{ip_v^2}{2|eE|(1-v^2)}]
{}_+\psi_{p^-}(x|\tilde A),\eqno(30)
$$
and similarly for other functions in (22). In other words, for $v\to1$ the
$\psi-$functions in (22) go over into $\psi_{p^-}$ up to an inessential
phase factor.

The functions $\psi_{p^-}$ for the potential (19) are defined as follows [5]
$$
{}^+\psi_{p^-}(x|\tilde A)={}^+\psi_{p^-}
=(4|eE|)^{-\frac14}\exp[-i\frac{p^-x^+}{2}-
ieE(\frac{x_3^2}{2}-\frac{(x^-)^2}{4})+\nu^*\ln\frac{\pi^-}{\sqrt{|eE|}}],
                                              \eqno(31)
$$
  $$
{}_+\psi_{p^-}=\theta(\pi^-)c_{1p}{}^+\psi_{p^-},\:{}^-\psi_{p^-}=
\theta(-\pi^-)c_{1p}\:{}_-\psi_{p^-,}\: \pi^-=p^--eEx^-,
$$
  $$
 \theta(x)=\left\{\begin{array}{cc}
1,\quad x>0\\
0,\quad x<0,
\end{array}\right.\quad  \nu^*=-\frac{i\lambda}{2}-\frac12,\quad
p^{\pm}=p^0\pm p_3, \;x^{\pm}=t\pm x_3.\eqno(32)
  $$
 For $\pi^-<0$ in (31) one must take $\pi^-=(-\pi^-)\exp[-i\pi]$. The
function ${}_-\psi_{p^-}$ is obtained from ${}^+\psi_{p^-}$ by changing sign
of $\pi^-$ under the logarithm sign:
$$
{}_-\psi_{p^-}(x|\tilde A)=(4|eE|)^{-\frac14}\exp[-i\frac{p^-x^+}{2}-
ieE(\frac{x_3^2}{2}-\frac{(x^-)^2}{4})+\nu^*\ln\frac{-\pi^-}{\sqrt{|eE|}}].
                                              \eqno(33)
$$
For $\pi^->0$ in (33) one must take $-\pi^-=\pi^-\exp[-i\pi]$.(For
$e=|e|$ see [7-8].)

The solutions $\psi_{p^-}$ are connected with the solutions with quantum
 number $p_3$ by the integral transformations [5]
$$
\varphi_{p_3}=\int_{-\infty}^{\infty}dp^-N(p_3,p^-)\psi_{p^-},\quad
N(p_3,p^-)=(2\pi|eE|)^{-\frac12}\exp\{-i\frac{(p^-)^2+4p^-p_3+2p_3^2}{4eE}\}.
                                                            \eqno(34)
$$
In these two expressions the charge $e$ may have any sign, but the
expressions for $\psi _{p^-}$ and $\varphi_{p_3}$ depend on the sign of $e$,
cf. [4, 7-8]. It is clear that (34) and similar expressions below are valid
in any gauge. The functions $\varphi_{p_3}$ differ from $\psi_{p_3}$ only
by the phase factor [5]
$$
\varphi_{p_3}=\exp[-\frac{i\lambda}{4}\ln2+i\frac{3\pi}{8}]\psi_{p_3}.\eqno(35)
$$
Taking into account the unitarity conditions for $N(p_3,p^-)$
$$
\int_{-\infty}^{\infty}dp^-N(p_3,p^-)N^*(p'_3,p^-)=\delta(p_3-p'_3),
$$
  $$
\int_{-\infty}^{\infty}dp_3N(p_3,p^-)N^*(p_3,p'{}^-)=\delta(p^--p'{}^-),
                                                              \eqno(36)
  $$
it is easy to express $\psi_{p^-}$ through $\varphi_{p_3}$
$$
\psi_{p^-}=\int_{-\infty}^{\infty}dp_3N^*(p_3,p^-)\varphi_{p_3}. \eqno(37)
$$
Using the same boost as in obtaining $\psi_{p_v}$ from $\psi_{p_3}$, we
 get from (34)
$$
\varphi_{p_v}=\int_{-\infty}^{\infty}dp^-N(p_v,p^-)\psi_{p^-},   \eqno(38)
$$
  $$
N(p_v,p^-)=(2\pi|eE|(1-v))^{-\frac12}\exp\{-i\frac{[p^-(1+v)]^2-4p^-p_v
(1+v)+2p_v^2}{4eE(1-v^2)}\}.                                     \eqno(39)
$$
This "matrix" is also unitary. Hence the reversed relation is
$$
\psi_{p^-}=\int_{-\infty}^{\infty}dp_vN^*(p_v,p^-)\varphi_{p_v}, \eqno(40)
$$

Combining (34) and (40) we find
$$
\varphi_{p_3}=\int_{-\infty}^{\infty}dp_vN(p_3,p_v)\varphi_{p_v}.
                                                               \eqno(41)
$$
  $$
N(p_3,p_v)= \int_{-\infty}^{\infty}dp^-N(p_3,p^-)N^*(p_v,p^-)=
  $$
   $$
(2\pi|eE|v))^{-\frac12}\exp\{i\frac{p_3^2(1-v^2)+2p_vp_3(1-v^2)
+p_v^2}{2veE(1-v^2)}-i\frac{\pi}{4}\}.                          \eqno(42)
   $$
 This matrix is also unitary , so
$$
\varphi_{p_v}=\int_{-\infty}^{\infty}dp_3N^*(p_3,p_v)\varphi_{p_3}.
                                                               \eqno(43)
$$

Now $\varphi_{p_3}$ satisfy the normalization condition [4]
$$
\int_{-\infty}^{\infty} dx_3{}_{\pm}\varphi_{p'_3}^*i\buildrel\leftrightarrow\over{\frac d{dt}}
{}_{\pm}\varphi_{p_3}=\pm2\pi\delta(p'_3-p_3)                  \eqno(44)
$$
and similarly for ${}^{\pm}\varphi_{p_3}$. Using (43) we have
$$
\int_{-\infty}^{\infty} dx_3{}_{\pm}\varphi_{p'_v}^
*i\buildrel\leftrightarrow\over{\frac d{dt}}
{}_{\pm}\varphi_{p_v}=
\int_{-\infty}^{\infty}dp_3\int_{-\infty}^{\infty}dp'_3
N(p'_3,p'_v)N^*(p_3,p_v)\int_{-\infty}^{\infty}
 dx_3{}_{\pm}\varphi_{p'_3}^*i\buildrel\leftrightarrow\over{\frac d{dt}}
{}_{\pm}\varphi_{p_3}.                                            \eqno(45)
$$
 Taking into account (44) and the unitarity of $N(p_3,p_v)$ we get (23),
 see (35).

We note also that relation (38) can be checked by direct calculation. So
inserting in it ${}^+\psi_{p^-}$ from (31) and using eq.(3.462.3) in [9],
we obtain
$$
{}^+\varphi_{p_v}=[2|eE|(1-v^2)]^{-\frac14}\exp\{-\frac{\pi\lambda}{8}+
i\frac{3\pi}{8}+i\frac{\lambda}{4}\ln\frac{1+v}{2(1-v)}+i\theta_v\}
D_{\nu^*}[(1+i)\tau_v]                                          \eqno(46)
$$
and similarly for other function of this set. Here $\theta_v$ is the same
as in (21). Comparison with (22) shows that $\varphi_{p_v}$ differs from
 $\psi_{p_v}$ only by a phase factor
$$
\varphi_{p_v}=\exp[i\frac{3\pi}{8}+
i\frac{\lambda}{4}\ln\frac{1+v}{2(1-v)}]\psi_{p_v}.              \eqno(47)
$$
As it should be the relation (41) goes over into (34) for $v\to1$. Really,
we can write $N(p_3,p_v)$ in (42) in the form
$$
N(p_3,p_v)=\tilde N(p_3,p_v)\exp[-\frac{i\pi}{4}+\frac{ip_v^2}{2|eE|(1-v^2)}],
\eqno(48)
$$
  $$
\tilde N(p_3,p_v)=(2\pi|eE|v))^{-\frac12}\exp\{i\frac{p_3^2(1+v)+2p_vp_3(1+v)
+p_v^2}{2v|eE|(1+v)}\}.                          \eqno(48a)
   $$
It is easy to see that
$$
\left.\tilde N(p_3,p_v)\right|_{v\to1}\to N(p_3,p^-),        \eqno(49)
$$
see (34). Besides, from (47) and (30) we have
$$
\left.\varphi_{p_v}\right|_{v\to1}\to\exp\{\frac{i\pi}{4} -\frac{ip_v^2}
{2|eE|(1-v^2)}\} \psi_{p^-}.                                    \eqno(50)
$$
So the phase factor on the right-hand side of (48) is cancelled by phase factor
on the right-hand side of (50).

Combining now  (43) and (41) we find
$$
\varphi_{p_{v'}}=\int_{-\infty}^{\infty}dp_vN(p_{v'},p_v)\varphi_{p_v},
$$
  $$
N(p_{v'},p_v)= \int_{-\infty}^{\infty}dp_3N^*(p_3,p_{v'})N(p_3,p_v)=
  $$
   $$
(-i(v'-v)2\pi|eE|)^{-\frac12}\exp\{
-i\frac{p_{v'}^2(1-vv')}{2|eE|(1-v'{}^2)(v'-v)}-i\frac{p_v^2(1-vv')}
{2|eE|(1-v^2)(v'-v)}+\frac{ip_{v'}p_v}{|eE|(v'-v)}\}.            \eqno(51)
   $$
 Here
$$
\sqrt{-i(v'-v)}=\left\{\begin{array}{cc}
e^{-i\frac{\pi}{4}}\sqrt{v'-v},\quad v'>v \\
 e^{i\frac{\pi}{4}}\sqrt{v-v'},\quad v'<v.
\end{array}\right.                                           \eqno(52)
 $$
$N(p_{v'},p_v)$ is hermitian and have group property
$$
N(p_{v''},p_v)= \int_{-\infty}^{\infty}dp_{v'}N(p_{v''},p_{v'})N(p_{v'},p_v).
                                                                \eqno(53)
  $$
If we insert into the right-hand side of (51) ${}^+\varphi_{p_v}=
{}^+\varphi_{p_v}(x|\tilde A)$ from (46) and use eq.(2.11.4.7) in [10]
we get the left-hand side of (51).

In this Section we have compared the solutions of Klein-Gordon equation
with vector-potential (19). Utilizing transformations similar to (19, 20) we
can go over to any vector-potential of the considered field.

\section{Stationary solutions}

We name "stationary" the solutions  with quantum number $p_s=p^0-sp_3$. For
$s=0$ these solutions are stationary in the usual sense. Others are
obtainable from these by boosts. All the consideration in this Section
is quite analogous to the one in the previous Section.

In the potential (10) we put
$$
a_{\mu}=\delta_{\mu0}a,\; k^{\mu}=(s\omega,0,0,\omega), \;
k^2=\omega^2(1-s^2),\; A^0=-A_0,\; a\omega=E,\; x_s=x_3-st.\eqno(54)
$$
From (14) and (13) we have
  $$
\tau_s=\frac{\pi_s}{\sqrt{|eE|(1-s^2)}},\quad \pi_s=p_s-|eE|x_s,\quad
p_s=p^0-sp_3.                                                     \eqno(55)
  $$
For the phase in (11) we get
$$
p_3x_3-p^0t-\frac{k\cdot p}{k^2}\varphi+\frac{ea\cdot k}{2k^2}\varphi^2=
-\frac{p_s(t-sx_3)}{1-s^2}+\frac{eEsx_s^2}{2(1-s^2)}.                \eqno(56)
$$
In this Section we bring the solutions with different $s$ to the potential
$$
{\cal A}_{\mu}=\delta_{\mu0}Ex_3, \quad {\cal A}^0=-{\cal A}_0. \eqno(57)
$$
In this gauge phase (56) acquires an additional term:
$$
\vartheta_s(x|{\cal A})\equiv\vartheta_s=-\frac{p_s(t-sx_3)}{1-s^2}+
\frac{eEsx_s^2}{2(1-s^2)}+\frac{eEst^2}2=
$$
 $$
  -\frac{p_s(t-sx_3)}{1-s^2}-\frac{|eE|s}{2(1-s^2)}(x_3^2-2stx_3+t^2). \eqno(58)
 $$

The classification in the family of sets $\psi_{p_s}$ is dictated by the
requirement that for $s=0$ we must obtain the classification in the set
$\psi_{p^0}$.( For some more details on classification in the latter
set  see [5]). So
$$
{}_{\pm}\psi_{p_s}=C_{ps}e^{i\vartheta_s}
D_{\pm i\frac{\lambda}{2}-\frac12}(\pm e^{\pm\frac{\pi}{4}}Z_s),\;
{}^{\pm}\psi_{p_s}=C_{ps}e^{i\vartheta_s}
D_{\mp i\frac{\lambda}{2}-\frac12}(\pm e^{\mp i\frac{\pi}{4}}Z_s),
$$
  $$
Z_s=-\sqrt2\tau_s=-\sqrt{\frac{2}{|eE|(1-s^2)}}(p_s-|eE|x_s),\;
C_{ps}=[2|eE|(1-s^2)]^{-\frac14}e^{-\frac{3\pi\lambda}{8}}.     \eqno(59)
$$
These functions satisfy the same relations (24) with the same $c_{1p}, c_{2p}$,
see (25).

Taking into account that the asymptotic expansions for the parabolic
 cylinder functions
contain $\exp[\pm i\tau_s^2/2]$-factors, we write down the expression
analogues to (29)
$$
\vartheta_s\pm\frac{\tau_s^2}{2}=\pm\frac{p_s^2}{2|eE|(1-s^2)}
\pm\frac{|eE|x_s^2}{2(1\pm s)}-\frac{p_sx^{\pm}}{1\pm s}-\frac{|eE|st^2}{2}.
                                                                   \eqno(60)
$$
For reasons mentioned after eq.(29) only the case with the upper sign is needed.
 Now it is
easy to verify that
$$
\left.{}^+\psi_{p_s}(x|{\cal A})\right|_{s\to1}\to\exp[-\frac{i3\pi}{8}-
\frac{i\lambda}{4}\ln\frac{2}{1-s^2}+\frac{ip_s^2}{2|eE|(1-s^2)}]
{}^+\psi_{p^-}(x|{\cal A}),                                      \eqno(61)
$$
  $$
{}^+\psi_{p^-}(x|{\cal A})=e^{ieEtx_3}\;{}^+\psi_{p^-}(x|{\tilde A}), \eqno(62)
$$
and similarly for other $\psi_{p_s}$. In other words, for
$s\to1$  $\psi_{p_s}$ go over into $\psi_{p^-}$ up to an inessential phase
factor. Now we note that instead potential (57) we can use (19). Then we
may say that the set $\psi_{p^-}$ constitute a bridge between $\psi_{p_v}$-
and $\psi_{p_s}$-families. All sets in this joint (super)family are
 indistinguishable experimentally. Instead of classical solution (1) we
have the superfamily of sets $(\psi_{p_v},\psi_{p_s})$ for the
Klein-Gordon equation.

As in previous Section, we consider $\varphi_{p_s}$ along with $\psi_{p_s}$
(cf. (38-39) and (47)):
$$
\varphi_{p_s}=\int_{-\infty}^{\infty}dp^-S(p_s,p^-)\psi_{p^-},\quad
 \psi_{p^-}=\int_{-\infty}^{\infty}dp_sS^*(p_s,p^-)\varphi_{p_s}.  \eqno(63)
$$
  $$
S(p_s,p^-)=(2\pi|eE|(1-s))^{-\frac12}\exp\{-i\frac{[p^-(1+s)]^2-4p^-p_s
(1+s)+2p_s^2}{4|eE|(1-s^2)},                                     \eqno(63a)
$$
  $$
\varphi_{p_s}=\exp[i\frac{\pi}{8}+
i\frac{\lambda}{4}\ln\frac{1+s}{2(1-s)}]\psi_{p_s}.              \eqno(64)
$$
 For $s=0$ we get from (63)
$$
\varphi_{p^0}=\int_{-\infty}^{\infty}dp^-S(p^0,p^-)\psi_{p^-},   \eqno(65)
$$
and due to unitarity of $S(p^0,p^-)$
$$
\psi_{p^-}=\int_{-\infty}^{\infty}dp^0S^*(p^0,p^-)\varphi_{p^0}. \eqno(66)
$$
We note that (65) was given in [5] and
  (63) can be obtained from (65) by going over to the boosted system.

Combining the first relation in (63) and (66) we find
$$
\varphi_{p_s}=\int_{-\infty}^{\infty}dp^0S^*(p^0,p_s)\varphi_{p^0},
                                                               \eqno(67)
$$
  $$
S^*(p^0,p_s)= \int_{-\infty}^{\infty}dp^-S(p_s,p^-)S^*(p^0,p^-)=
  $$
   $$
(2\pi|eE|s))^{-\frac12}\exp\{i\frac{p_0^2(1-s^2)-2p_sp^0(1-s^2)
+p_s^2}{2s|eE|(1-s^2)}-i\frac{\pi}{4}\}.                          \eqno(68)
   $$
From (64) and (61) we have
$$
\left.\varphi_{p_s}\right|_{s\to1}\to\exp\{-\frac{i\pi}{4} +\frac{ip_s^2}
{2|eE|(1-s^2)}\} \psi_{p^-},                                    \eqno(69)
$$
and (68) can be written as
$$
S^*(p^0,p_s)=\tilde S^*(p^0,p_s)\exp[-\frac{i\pi}{4}+\frac{ip_s^2}
{2|eE|(1-s^2)}],
$$
  $$
\tilde S^*(p^0,p_s)=(2\pi|eE|s))^{-\frac12}\exp\{i\frac{p_0^2(1+s)-2p_sp^0(1+s)
+p_s^2}{2s|eE|(1+s)}\}.                                      \eqno(70)
   $$
Now
$$
\left.\tilde S^*(p^0,p_s)\right|_{s\to1}\to S^*(p^0,p^-)=(2\pi|eE|)^{-\frac12}
\exp\{\frac{i(p^-)^2}{4|eE|}-\frac{ip^0p^-}{|eE|}+\frac{ip_0^2}{2|eE|}\}.
        \eqno(71)
$$
From (69-71) it follows that for $s\to1$ (67) goes over to (66).

As shown in [5], $\psi_{p^0}$ (i.e. functions in (59) for $s=0$) are
orthonormalized. The same is true for $\psi_{p_s}$ in (52) due to unitarity
of $S(p^0,p_s)$, cf text near eqs. (44-45).

The reversed relation for (67) has the form
$$
\varphi_{p^0}=\int_{-\infty}^{\infty}dp_sS(p^0,p_s)\varphi_{p_s}.
                                                               \eqno(72)
$$
Combining (67) and (72) we find
$$
\varphi_{p_{s'}}=\int_{-\infty}^{\infty}dp_sS(p_{s'},p_s)\varphi_{p_s},
$$
  $$
S(p_{s'},p_s)= \int_{-\infty}^{\infty}dp^0S^*(p^0,p_{s'})S(p^0,p_s)=
  $$
   $$
(i(s'-s)2\pi|eE|)^{-\frac12}\exp\{
i\frac{p_{s'}^2(1-ss')}{2|eE|(1-s'{}^2)(s'-s)}+i\frac{p_s^2(1-ss')}
{2|eE|(1-s^2)(s'-s)}-\frac{ip_{s'}p_s}{|eE|(s'-s)}\}            \eqno(73)
   $$
 Here
$$
\sqrt{i(s'-s)}=\left\{\begin{array}{cc}
e^{i\frac{\pi}{4}}\sqrt{s'-s},\quad s'>s \\
 e^{-i\frac{\pi}{4}}\sqrt{s-s'},\quad s'<s.
\end{array}\right.                                           \eqno(74)
 $$
$S(p_{s'},p_s)$ is hermitian and have group property
$$
S(p_{s''},p_s)= \int_{-\infty}^{\infty}dp_{s'}S(p_{s''},p_{s'})S(p_{s'},p_s).
                                                                \eqno(75)
  $$

Now combining the first relation in (63) and (40) we obtain
$$
\varphi_{p_s}=\int_{-\infty}^{\infty}dp_vI(p_s,p_v)\varphi_{p_v},
                                                               \eqno(76)
$$
  $$
I(p_s,p_v)= \int_{-\infty}^{\infty}dp^-S(p_s,p^-)N^*(p_v,p^-)=
  $$
   $$
(\pi|eE|f_+))^{-\frac12}\exp\{\frac{if_-}{2|eE|f_+}[\frac{p_s^2}{1-s^2}-
\frac{p_v^2}{1-v^2}]+\frac{2ip_sp_v}{|eE|f_+}-\frac{i\pi}4\},
\; f_{\pm}=(1+s)(1-v)\pm(1-s)(1+v).                    \eqno (77)
   $$
Similarly, from (38) and (63) we find
$$
\varphi_{p_v}=\int_{-\infty}^{\infty}dp_sI^*(p_s,p_v)\varphi_{p_s},
                                                               \eqno(78)
$$
For $s=v=0$ we get from (76)
$$
\varphi_{p^0}=\int_{-\infty}^{\infty}dp_3I(p^0,-p_3)\varphi_{p_3}
                                                               \eqno(79)
$$
in agreement with eq. (117) in [5].

If we use in the right-hand side of (76) the expression for
${}^+\varphi_{p_v}(x|{\cal A})=e^{-i|eE|tx_3}\;{}^+\varphi_{p_v}$,
(where ${}^+\varphi_{p_v}$ is defined in (47), (22), and (21)) and utilize
formula (2.11.4(7)) in [5], we get ${}^+\varphi_{p_s}(x|{\cal A})$ defined in
 (64), (59), and (58).
\section{Conclusion}

Any set in the collection of sets($\psi_{p_v},\psi_{p_s}$) can be obtained
from any other by gauge manipulations. The classification in any set is
dictated by the classification in the set $\psi_{p_3}$, where it is beyond
doubt.
\section{Acknowledgments}
This work was supported in part by the Russian Foundation for Basic Research
(projects no 00-15-96566 and 01-02-30024).

 \section*{References}
\begin{enumerate}
\item
  A.Hansen, F.Ravndal, Physica Scripta, {\bf 23}, 1036 (1981).      \\
\bibitem{2}
  W.Greiner, B.M\"uller, J.Rafelski, {\sl Quantum Electrodynamics of Strong
    Field}, Springer-Verlag (1985). \\
\bibitem{3}
  A.Calogeracos, N.Dombey, Contemp. Phys. {\bf 40}, 313 (1999).         \\
\item
 A.I.Nikishov, Tr. Fiz. Inst. Akad. Nauk SSSR {\bf 111}, 152 (1979)  ;\\
J. Sov. Laser Res. {\bf 6}, 619 (1985).
\bibitem{5}
A.I.Nikishov, hep-th/0111137.
\bibitem{6}
  {\sl Higher Transcendental Functions, Vol. 2 (Bateman Manuscript Project
     )}, Ed. by  A.Erd\'elyi (McGraw-Hill, New York, 1953; Nauka, Moscow,
    1980; Pergamon, Oxford, 1982). \\
\bibitem{7}
 N.B.Narozhny and A.I.Nikishov, Teor.  Mat. Fiz. {\bf 26}, 16 (1976).  \\
\item
 N.B.Narozhny and A.I.Nikishov, Tr. Fiz. Inst. Akad. Nauk SSSR
 {\bf 168}, 175 (1985); \\
 in {\sl Issues in Intensive-Field Quantum Electrodynamics}, Ed. by
 V.L.Ginzburg (Nova Science, Commack, 1987).   \\

\bibitem{9}
 I.S.Gradstein, I.M.Ryzhik, {\sl Tables of Integrals, Sums, Series,
    and Products }, Moscow, 1962.                    \\
\bibitem{10}
 A.P.Prudnikov, Yu. A.Brychkov, and O.I.Marichev, {\sl Integrals and
    Series. Special Functions}, Moscow, Nauka, 1983.
 \end{enumerate}
\end{document}